\documentclass[prd,showpacs,showkeys,nofootinbib,preprint]{revtex4-1}
\usepackage{amsmath,amsfonts,amssymb,color}
\usepackage{bm}
\usepackage{pgfplots}
\usepackage[colorlinks=true,urlcolor=blue,linkcolor=blue,citecolor=blue]{hyperref}

\begin{document}

\title{Parity violation in Poincar\'e gauge gravity}

\author{Yuri N. Obukhov}

\affiliation{Russian Academy of Sciences, Nuclear Safety Institute, 
B.Tulskaya 52, 115191 Moscow, Russia
\email{obukhov@ibrae.ac.ru}}

\begin{abstract}
We analyse the parity violation issue in the Poincar\'e gauge theory of gravity for the two classes of models which are built as natural extensions of the Einstein-Cartan theory. The conservation laws of the matter currents are revisited and we clarify the derivation of the effective Einstein field equation and the structure of the effective energy-momentum current for arbitrary matter sources.
\end{abstract}

\maketitle

\section{Introduction}

Experimental discovery of parity $P$ and $CP$ violation in the weak interaction processes \cite{CP} strongly affected the field-theoretic developments of the high energy physics and essentially contributed to the establishment of the standard model of electroweak interactions. A natural question is then are there additional parity violation effects other than those in electroweak interactions, and in particular could the parity violation be manifest in gravitational interactions? In the absence of convincing theoretical or experimental arguments which could rule out the violation of parity in gravity, one may ask: is gravity parity violating, and if yes, what are the possible physical effects?

Presently, the interest to the parity violating sector of physical interactions \cite{OPZ,Chen,Ho2,Ho3,Diakonov,Baekler1,Baekler2,Capo,Karananas,BC,Iosifidis} is apparently growing. The possibility of extending the gravitational Lagrangian by parity odd terms has been proposed already in the 1960's \cite{LO}, and such extensions were later investigated \cite{Hari,Hoj,MH,Purcell,Mukh} both in classical and quantum domains, including Ashtekar's canonical formalism and loop quantum gravity \cite{Holst,CK}. The discussion of parity-nonconserving terms looks important for understanding of the baryon asymmetry of the universe \cite{ADS}, where these terms can emerge from the structure of the quantum vacuum \cite{Randono,Pop,Bj}. The search for electric dipole moments of fundamental particles and the study of related effects in inertial and gravitational fields represents another important issue \cite{POS,ostor}.

A considerable attention in the current literature was attracted to the study of the so called Chern-Simons (CS) modified gravity theories \cite{Alex:2008d,Alex:2008b,Xu}. It is worthwhile to notice that the basic features of the classical Einstein's general relativity theory (GR) are preserved in CS modified gravity, while the naturally arising parity violation leads to a number of physically interesting effects. One of the key signatures of parity violation is expected for gravitational waves \cite{Alex:2018,Nishizawa,Conroy,Yoshida,Qiao,Zhao}, known as the amplitude birefringence in their propagation. Another possible manifestation of parity violation could be the loss of power of inflationary perturbations resulting in a certain suppression of parity-odd modes in the power spectrum of the quantum fluctuations responsible for the subsequent formation of the large scale structure of the universe, with possible observational effects in the form of anisotropy of the WMAP data for low multipoles \cite{Alex:2014}. Qualitatively, CS modified gravity theories can be considered as extensions of the axion models which predict similar cosmological and astrophysical implications \cite{Carroll,lue,Jackiw,Castillo:2015}.

Our attention is here focused on the parity violating gravitational models in the framework of the Poincar\'e gauge theory of gravity (PG) which is based on the Riemann-Cartan spacetime geometry with the curvature and the torsion. Citing Freund: ``Parity violation and torsion go hand-in-hand'' \cite{Freund}. The parity violating terms in the gravitational Lagrangian have a close relation to the axial anomaly in the Riemann-Cartan spaces \cite{Obukhov:1982,Obukhov:1983,Obukhov:1997,Yajima:1985,Yajima:1996,Chandia}, which revealed an important role of the Nieh-Yan topological invariant \cite{Nieh:1982,Nieh:2018,Hehl:1991,Li:1999}. Interesting applications of the corresponding formalism include the study of the chiral torsional effect \cite{Imaki,Khaidukov,Nissinen,Volovik} in condensed matter systems and of the vortical effects in the heavy-ion collisions \cite{ion1,ion2,ion4,ion5}.

The gauge approach to gravity was started by the early works of Weyl, Cartan, Fock, continued by Utiyama, and finally developed by Sciama \cite{Sciama} and Kibble \cite{Kibble} to establish the Einstein-Cartan model \cite{Trautman:2006} as a viable Poincar\'e gravity theory. Historic and technical overview of the gauge gravity can be found in \cite{Hehl:1976,Shapiro,Blagojevic,selected,overview,Ponomarev}, whereas the recent book \cite{Reader} contains an exhaustive bibliography. When applied to the 10-parameter Poincar\'e group $G\!=\!T_4\!\rtimes\!SO(1,3)$, the Yang-Mills-Sciama-Kibble scheme identifies the 10-plet of the gauge potentials with the coframe $\vartheta^\alpha = e^\alpha_i dx^a$ (corresponding to the translation subgroup $T_4$) and the local connection $\Gamma^{\alpha\beta} = -\,\Gamma^{\beta\alpha} = \Gamma_{i}{}^{\alpha\beta} dx^i$ (for the Lorentz subgroup $SO(1,3)$). The ``translational'' and ``rotational'' Poincar\'e gauge field strengths $T^\alpha$ and $R^{\alpha\beta}$ are then interpreted as the torsion (\ref{Tor}) and the curvature (\ref{Cur}) 2-forms, thus naturally introducing the Riemann-Cartan geometry \cite{LdB} on the spacetime manifold.

Our basic notation and conventions are as follows \cite{MAG}: Greek indices $\alpha, \beta,\dots{} = 0, \dots, 3$, denote the anholonomic components (for example, of a coframe $\vartheta^\alpha$), while the Latin indices $i,j,\dots{} =0,\dots, 3$, label the holonomic components ($dx^i$, e.g.). From the volume 4-form $\eta$, the $\eta$-basis is constructed with the help of the interior products as $\eta_{\alpha_1 \dots\alpha_p}:= e_{\alpha_p}\rfloor\dots e_{\alpha_1}\rfloor\eta$, $p=1,\dots,4$. These forms are related to the $\vartheta$-basis via the Hodge dual operator $^\ast$, for example, $\eta_\alpha = {}^\ast\vartheta_\alpha$ and $\eta_{\alpha\beta} = {}^\ast\left(\vartheta_\alpha\wedge\vartheta_\beta\right)$. For skew-symmetric objects $\varphi_{\alpha\beta} = -\,\varphi_{\beta\alpha}$ (forms of any rank), we define the right dual $\varphi^\star_{\alpha\beta} = {\frac 12}\eta_{\alpha\beta\mu\nu}\varphi^{\mu\nu}$ which is different from Hodge star $^\ast$. The Minkowski metric reads $g_{\alpha\beta} = {\rm diag}(+1,-1,-1,-1)$.

\section{Matter sources of the gravitational gauge field}

The gravitational PG fields are coupled to the Noether currents of the Poincar\'e group: the energy-momentum ${\mathfrak T}_\alpha$ and the spin ${\mathfrak S}_{\alpha\beta} = -\,{\mathfrak S}_{\beta\alpha}$ of matter \cite{Hehl:2020}.

Quite generally, one can describe matter by a tensor-valued $p$-form $\psi^A$ (tensor structure is encoded in the multi-index \begin{footnotesize}$A$\end{footnotesize}), which transforms under the local action of the Lorentz group as
\begin{equation}\label{lor2}
\delta\vartheta^{\alpha} = \varepsilon_{\beta}{}^{\alpha}\,\vartheta^{\beta},\qquad 
\delta\Gamma^{\alpha\beta} = - D\varepsilon^{\alpha\beta},\qquad \delta
\psi^A = -\,{\frac 12}\varepsilon^{\alpha\beta}\,(\rho_{\alpha\beta})^A{}_B\,\psi^B.
\end{equation}
Here $\varepsilon^{\alpha\beta} = -\,\varepsilon^{\beta\alpha}$ are arbitrary local parameters, $(\rho_{\alpha\beta})^A{}_B = -\,(\rho_{\alpha\beta})^A{}_B$ are the $SO(1,3)$ generators, and $D\varepsilon^{\alpha\beta} = d\varepsilon^{\alpha\beta} + \Gamma_\gamma{}^\alpha\varepsilon^{\gamma\beta} + \Gamma_\gamma{}^\beta\varepsilon^{\alpha\gamma}$. The covariant derivative of the matter field is defined by $D\psi^A = d\psi^A - {\frac 12}\Gamma^{\alpha\beta}\wedge(\rho_{\alpha\beta})^A{}_B\psi^B$. 

The dynamics of matter is determined by the action principle for a general Lagrangian 4-form (allowing for the nonminimal coupling)
\begin{eqnarray} 
L = L(\psi^A, D\psi^A, \vartheta^\alpha, T^\alpha, R^{\alpha\beta})\,,\label{L} 
\end{eqnarray}
and the matter field equations are derived from the variation with respect to $\psi^A$:
\begin{equation}
{\frac {\delta L} {\delta\psi^A}} = {\frac {\partial  L}{\partial\psi^A}} - (-1)^{p}D\,{\frac {\partial L}{\partial D\psi^A}} = 0.\label{dLdpsi}
\end{equation}
The matter sources of the gravitational field are 3-forms of the energy-momentum and the spin currents, respectively,
\begin{eqnarray}
{\mathfrak T}_{\alpha} &:=& -\,{\frac {\delta L}{\delta\vartheta^{\alpha}}} =  
- \,{\frac {\partial L}{\partial\vartheta^{\alpha}}} 
- D\,{\frac {\partial L}{\partial T^{\alpha}}}\, ,\label{sigC0}\\
c{\mathfrak S}_{\alpha\beta} &:=& -\,2{\frac {\delta L}{\delta\Gamma^{\alpha\beta}}} =   
(\rho_{\alpha\beta})^A{}_B\psi^B\wedge{\frac {\partial L} {\partial D\psi^A}}
- 2\vartheta_{[\alpha}\wedge {\frac {\partial L}{\partial T^{\beta]}}} 
-2D{\frac {\partial L}{\partial R^{\alpha\beta}}}\,.\label{spin0}
\end{eqnarray}

In accordance with the Noether theorem, the invariance of $L$ under the local diffeomorphisms on the spacetime manifold and under the local Lorentz group yields the energy-momentum and the angular momentum conservation laws
\begin{eqnarray}\label{conmomC}
D{\mathfrak T}_\alpha = (e_\alpha\rfloor T^\beta)\wedge{\mathfrak T}_\beta
+ {\frac 12}(e_\alpha\rfloor R^{\beta\gamma})\wedge c{\mathfrak S}_{\beta\gamma},\\
cD{\mathfrak S}_{\alpha\beta} + \vartheta_\alpha\wedge
{\mathfrak T}_\beta - \vartheta_\beta\wedge{\mathfrak T}_\alpha = 0,\label{Noe2}
\end{eqnarray}
when the matter field $\psi^A$ satisfies the equations of motion (\ref{dLdpsi}). The Noether theorem also fixes the explicit form of the canonical energy--momentum current:
\begin{eqnarray}
{\mathfrak T}_\alpha &=&  (e_\alpha\rfloor D\psi^A)\wedge {\frac{\partial L}{\partial D\psi^A}}
+ (e_\alpha\rfloor\psi^A)\wedge{\frac{\partial L}{\partial\psi^A}} - e_\alpha\rfloor L\nonumber\\
&& -\,D{\frac{\partial L}{\partial T^\alpha}} + (e_{\alpha}\rfloor T^\beta)\wedge
{\frac{\partial L}{\partial T^\beta}} + (e_{\alpha}\rfloor R^{\beta\gamma})\wedge 
{\frac{\partial L}{\partial R^{\beta\gamma}}}.\label{momC}
\end{eqnarray}

Both macro- and microscopic matter sources are important in PG theory. 

A physically viable description of {\it macroscopic matter} with spin is provided by Weyssenhoff's fluid \cite{Weyss,OK} model. The latter represents a special case of a classical continuous medium with microstructure, which is characterized by the flow 3-form $u = u^\alpha\eta_\alpha$, the spin density ${\mathcal S}_{\alpha\beta} = -\,{\mathcal S}_{\beta\alpha}$, the particle density $\rho$, the internal energy density $\varepsilon$ and the fluid pressure $p$. The covariant spin density satisfies the Frenkel condition $u^\beta{\mathcal S}_{\alpha\beta} = 0$ and its dynamics is governed by the equation of motion
\begin{equation}
D(u\,{\mathcal S}_{\alpha\beta}) - {\frac 1{c^2}}u_\beta u^\gamma D(u\,{\mathcal S}_{\alpha\gamma})
- {\frac 1{c^2}}u_\alpha u^\gamma D(u\,{\mathcal S}_{\gamma\beta}) = 0.\label{spineq}
\end{equation}

The corresponding matter sources (energy-momentum and spin currents) read 
\begin{eqnarray}\label{TW}
{\mathfrak T}_{\alpha} &=& {\frac 1{c^2}}\Big[\varepsilon u_\alpha u - {}^\ast\!D
(u\,{\mathcal S}_{\alpha\beta})\,u^\beta u\Big] - p\Big(\eta_\alpha - {\frac 1{c^2}}u_\alpha u\Big),\\
c{\mathfrak S}_{\alpha\beta} &=& u\,{\mathcal S}_{\alpha\beta}.\label{SW} 
\end{eqnarray}

As for the {\it microscopic matter}, we describe it in terms of the Dirac spin ${\frac 12}$ field $\Psi$. In the Clifford algebra-valued exterior calculus, the basic object is the matrix-valued 1-form $\gamma = \gamma_\alpha\,\vartheta^\alpha$. Unlike the usual forms, such objects do not anticommute; in particular, ${\frac i{4!}}\gamma\wedge\gamma\wedge\gamma\wedge\gamma = \gamma_5\eta$, and $i\gamma\wedge\gamma = \sigma_{\alpha\beta}\,\vartheta^\alpha\wedge\vartheta^\beta$, with $\sigma_{\alpha\beta} = i\gamma_{[\alpha}\gamma_{\beta]}$.

The dynamics of a relativistic fermion particle with mass $m$ is described by Dirac wave equation
spinor field:
\begin{eqnarray}
i\hbar{}^\ast\gamma\wedge \left(D\,\Psi - \hbox{$\scriptstyle\frac{1}{2}$}
T\,\Psi\right) + {}^\ast mc\,\Psi = 0,\label{dirRCa}
\end{eqnarray}
with the spinor covariant derivative $D\Psi = d\Psi + \frac{i}{4}\Gamma^{\alpha\beta}\wedge\sigma_{\alpha\beta}\,\Psi$.

For the canonical energy-momentum and spin currents (\ref{momC}) and (\ref{spin0}) we find
\begin{eqnarray}\label{Sa}
{\mathfrak T}_\alpha &=& {\frac {i\hbar c}{2}}\left(\overline{\Psi}\,{}^\ast\!\gamma 
D_\alpha\Psi - D_\alpha\overline{\Psi}\,{}^\ast\!\gamma\Psi\right),\\
{\mathfrak S}_{\alpha\beta} &=& {\frac {\hbar}{2}}\,\vartheta_\alpha\wedge\vartheta_\beta
\wedge\overline{\Psi}\gamma\gamma_5\Psi.\label{Tau}
\end{eqnarray}
Hereafter $D_\alpha:=e_\alpha\rfloor D$, and the Dirac-conjugate spinors are denoted by $\overline{\Psi}$.

\section{Extended Einstein-Cartan model}

Let us consider a generalization of the Hilbert-Einstein Lagrangian
\begin{equation}
V_{HE} = {\frac {1}{2\kappa c}}\left(\eta_{\alpha\beta} + \overline{a}{}_0
\vartheta_\alpha\wedge\vartheta_\beta\right)\wedge R^{\alpha\beta}.\label{LH}
\end{equation}
Here $\kappa = {\frac {8\pi G}{c^4}}$ is Einstein's gravitational constant. The dimensionless coupling constant $\overline{a}{}_0$ is responsible for the parity violating effects (note that $\xi = {\frac 1{\overline{a}{}_0}}$ is often called a Barbero-Immirzi parameter). The extended Einstein-Cartan model (\ref{LH}) was originally proposed by Hojman, Mukku and Sayed \cite{Hoj} and later revisited by Holst \cite{Holst} in the framework of Ashtekar's approach to the canonical quantum gravity. The parity violating effects in the Einstein-Cartan model were analysed earlier in \cite{Freidel,Mercuri:2008,Mercuri:2009,Mercuri:2010,Kazmierczak}.

The Einstein-Cartan field equations are derived from the variation of the total Lagrangian
\begin{equation}
V_{HE}(\vartheta^{\alpha}, T^{\alpha}, R^{\alpha\beta}) + {\frac 1c}
L(\psi^A, D\psi^A, \vartheta^{\alpha}, T^{\alpha}, R^{\alpha\beta})\label{Ltot}
\end{equation}
with respect to the coframe and the local Lorentz connection, and read explicitly
\begin{eqnarray}\label{ECH1}
{\frac 12}\eta_{\alpha\beta\gamma}\wedge R^{\beta\gamma} + \overline{a}{}_0R_{\alpha\beta}
\wedge\vartheta^\beta &=& \kappa\,{\mathfrak T}_\alpha,\\
\eta_{\alpha\beta\gamma}\wedge T^{\gamma} + \overline{a}{}_0\left(T_\alpha\wedge\vartheta_\beta
- T_\beta\wedge\vartheta_\alpha \right) &=&\kappa c\,{\mathfrak S}_{\alpha\beta}.\label{ECH2}
\end{eqnarray}
By expanding the currents with respect to the $\eta$-basis, we obtain the energy-momentum tensor and the spin density tensor: ${\mathfrak T}_\alpha = {\mathfrak T}_\alpha{}^\mu\eta_\mu$, and ${\mathfrak S}_{\alpha\beta} = {\mathfrak S}_{\alpha\beta}{}^\mu\eta_\mu$. Substituting $R^{\alpha\beta} = {\frac 12}R_{\mu\nu}{}^{\alpha\beta}\,\vartheta^\mu\wedge\vartheta^\nu$ and $T^{\alpha} = {\frac 12}T_{\mu\nu}{}^\alpha\,\vartheta^\mu\wedge\vartheta^\nu$, we find the Einstein-Cartan field equations in components
\begin{eqnarray}\label{EC1}
{\rm Ric}_\alpha{}^\beta - {\frac 12}\delta_\alpha^\beta\,R - \overline{a}{}_0\,R_{\mu\nu\lambda\alpha}\,\eta^{\mu\nu\lambda\beta} 
&=& \kappa\,{\mathfrak T}_\alpha{}^\beta,\\
T_{\alpha\beta}{}^\gamma - \delta_\alpha^\gamma T_{\lambda\beta}{}^\lambda + \delta_\beta^\gamma
T_{\lambda\alpha}{}^\lambda - {\frac {\overline{a}{}_0}{2}}\,\eta_{\alpha\beta}{}^{\mu\nu}
(T_{\mu\nu}{}^\gamma - \delta_\mu^\gamma T_{\lambda\nu}{}^\lambda + \delta_\nu^\gamma T_{\lambda\mu}{}^\lambda)
&=& \kappa c\,{\mathfrak S}_{\alpha\beta}{}^\gamma.\label{EC2}
\end{eqnarray}

Since the second field equation (\ref{ECH2}) is algebraic, one can resolve it and express the spacetime torsion in terms of the spin of matter. Taking the right dual of (\ref{ECH2}), which technically means a contraction with the Levi-Civita tensor, we find
\begin{equation}
-\,\left(T_\alpha\wedge\vartheta_\beta - T_\beta\wedge\vartheta_\alpha \right) 
+ \overline{a}{}_0\eta_{\alpha\beta\gamma}\wedge T^{\gamma} = \kappa c\,
{\mathfrak S}^\star_{\alpha\beta}.\label{ECH3}
\end{equation}
The algebraic system (\ref{ECH2}) and (\ref{ECH3}) can be solved with respect to $\eta_{\alpha\beta\gamma}\wedge T^{\gamma}$ and $(T_\alpha\wedge\vartheta_\beta - T_\beta\wedge\vartheta_\alpha)$. This yields
\begin{equation}
T_\beta\wedge\vartheta_\alpha - T_\alpha\wedge\vartheta_\beta = \zeta\kappa c\,
\Big({\mathfrak S}^\star_{\alpha\beta} - \overline{a}{}_0{\mathfrak S}_{\alpha\beta}\Big).\label{ECH4}
\end{equation}
Here we introduced a new parameter
\begin{equation}
\zeta = {\frac {1}{1 + \overline{a}{}_0^2}} = {\frac {\xi^2}{1 + \xi^2}}.\label{zeta0}
\end{equation}
We now apply the technique of ($\varphi$-$\psi$-$\chi$)-maps (see \ref{math}), and in particular use (\ref{map23}) to write the spin 3-form as 
\begin{equation}
{\mathfrak S}_{\beta\gamma} = \vartheta_\beta\wedge\{\psi({\mathfrak S})\}_\gamma 
- \vartheta_\gamma\wedge\{\psi({\mathfrak S})\}{}_\beta,\label{Spsi}
\end{equation}
and use a similar representation for ${\mathfrak S}^\star_{\alpha\beta}$. As a result, from (\ref{ECH4}) we derive the torsion 2-form as a function of the spin:
\begin{equation}
T_\alpha = \zeta\kappa c\Big(\left\{\psi({\mathfrak S}^\star)\right\}_\alpha - \overline{a}{}_0
\left\{\psi({\mathfrak S})\right\}_\alpha\Big).\label{Tsol}
\end{equation}
Furthermore, recalling (\ref{TK}) and using the map (\ref{map12}), we compute the contortion 1-form
\begin{equation}
K_{\alpha\beta} = \zeta\kappa c\Big(\left\{\chi({\mathfrak S}^\star)\right\}_{\alpha\beta}
- \overline{a}{}_0\left\{\chi({\mathfrak S})\right\}_{\alpha\beta}\Big),\label{K1}
\end{equation}

After we obtained the torsion explicitly (\ref{Tsol}) and (\ref{K1}), we can recast the first field equation (\ref{ECH1}) into the form of the effective Einstein equation. With the help of (\ref{RR}), we derive
\begin{eqnarray}
{\frac 12}\eta_{\alpha\beta\gamma}\wedge R^{\beta\gamma} = {\frac 12}\eta_{\alpha\beta\gamma}
\wedge \widetilde{R}^{\beta\gamma} - {\frac 12}\eta_{\alpha\beta\gamma}\widetilde{D}
K^{\beta\gamma} + {\frac 12}\eta_{\alpha\beta\gamma}\wedge K_\lambda{}^\gamma\wedge 
K^{\beta\lambda}.\label{etaR}
\end{eqnarray}
Transform the second term on the right-hand side:
\begin{eqnarray}
- \,{\frac 12}\eta_{\alpha\beta\gamma}\widetilde{D} K^{\beta\gamma} = \widetilde{D}\Big(
{\frac 12}\eta_{\alpha\beta\gamma}\wedge K^{\beta\gamma}\Big) 
= -\,\widetilde{D}\left(K^\star_{\alpha\lambda}\wedge\vartheta^\lambda\right).\label{DK}
\end{eqnarray}
For the last term in (\ref{etaR}), 
\begin{equation}
{\frac 12}\eta_{\alpha\beta\gamma}\wedge K_\lambda{}^\gamma\wedge K^{\beta\lambda} = {\frac 12}
\eta_{\alpha\beta\gamma\mu}\,\vartheta^\mu\wedge K_\lambda{}^\gamma\wedge K^{\beta\lambda},
\end{equation}
we use the identity which holds in 4 dimensions 
\begin{equation}
\eta_{\alpha\beta\gamma\mu}\,K_\lambda{}^\gamma \equiv \eta_{\lambda\beta\gamma\mu}\,K_\alpha{}^\gamma 
+ \eta_{\alpha\lambda\gamma\mu}\,K_\beta{}^\gamma + \eta_{\alpha\beta\lambda\mu}\,K_\gamma{}^\gamma
+ \eta_{\alpha\beta\gamma\lambda}\,K_\mu{}^\gamma.
\end{equation}
Noticing that $K_\gamma{}^\gamma = 0$, we reshuffle the rest of the terms to derive 
\begin{equation}
{\frac 12}\eta_{\alpha\beta\gamma}\wedge K_\lambda{}^\gamma\wedge K^{\beta\lambda} = -\,{\frac 12}
\left(K^\star_{\alpha\beta}\wedge K^\beta{}_\gamma + K_{\alpha\beta}\wedge K^\star{}^\beta{}_\gamma
\right)\wedge\vartheta^\gamma.\label{eKK}
\end{equation}
With an account of (\ref{DK}) and (\ref{eKK}) we thus identically recast (\ref{etaR}) into
\begin{equation}
{\frac 12}\eta_{\alpha\beta\gamma}\wedge R^{\beta\gamma} = {\frac 12}\eta_{\alpha\beta\gamma}\wedge
\widetilde{R}^{\beta\gamma} - \widetilde{D}(K^\star_{\alpha\lambda}\wedge\vartheta^\lambda)
-\,{\frac 12}\,(K^\star_{\alpha\beta}\wedge K^\beta{}_\lambda + K_{\alpha\beta}\wedge 
K^\star{}^\beta{}_\lambda)\wedge\vartheta^\lambda.\label{etaR2}
\end{equation}
Next, again making use of the curvature decomposition (\ref{RR}), we derive
\begin{equation}\label{Rth}
R_{\alpha\beta}\wedge\vartheta^\beta = -\,\widetilde{D}\left(K_{\alpha\lambda}\wedge
\vartheta^\lambda\right) - K_{\alpha\beta}\wedge K^\beta{}_\lambda\wedge\vartheta^\lambda,
\end{equation}
where we took into account the Ricci identity $\widetilde{R}_{\alpha\beta}\wedge\vartheta^\beta\equiv 0$. 

Finally, inserting (\ref{etaR2}) and (\ref{Rth}) into (\ref{ECH1}), we recast the latter into an effective Einstein equation
\begin{equation}
{\frac 12}\eta_{\alpha\beta\gamma}\wedge \widetilde{R}^{\beta\gamma} = \kappa
\,{\mathfrak T}^{\rm eff}_\alpha,\label{Eeff}
\end{equation}
with the effective energy-momentum current
\begin{eqnarray}
{\mathfrak T}^{\rm eff}_\alpha = {\mathfrak T}_\alpha + {\frac 1\kappa}\widetilde{D}
\left(\widehat{K}_{\alpha\lambda}\wedge\vartheta^\lambda\right) 
+ \,{\frac 1{2\kappa}}\left(\widehat{K}_{\alpha\beta}\wedge K^\beta{}_\lambda 
+ K_{\alpha\beta}\wedge \widehat{K}^\beta{}_\lambda\right)\wedge\vartheta^\lambda,\label{Teff1}
\end{eqnarray}
where we introduced
\begin{equation}
\widehat{K}_{\alpha\beta} = K^\star_{\alpha\beta} + \overline{a}{}_0K_{\alpha\beta}.\label{Khat}
\end{equation}
Recalling (\ref{K1}), we have for the right dual
\begin{equation}
K^\star_{\alpha\beta} = \zeta\kappa c\Big(\left\{\chi^\star({\mathfrak S}^\star)\right\}_{\alpha\beta}
- \overline{a}{}_0\left\{\chi^\star({\mathfrak S})\right\}_{\alpha\beta}\Big).\label{K2}
\end{equation}
Then making use of the identities (\ref{cc12}) we easily derive for (\ref{Khat}):
\begin{equation}\label{Khat2}
\widehat{K}_{\alpha\beta} = -\,\kappa c\left\{\chi({\mathfrak S})\right\}_{\alpha\beta}.
\end{equation}
In particular, recalling (\ref{map12}), this yields
\begin{equation}\label{Kpsi}
\widehat{K}_{\alpha\beta}\wedge\vartheta^\beta = -\,\kappa c \{\psi({\mathfrak S})\}_\alpha.
\end{equation}

\section{Conservation laws in Einstein-Cartan model}

Let us analyse the conservation laws of the total angular momentum (\ref{Noe2}) and of the energy-momentum (\ref{conmomC}) in the framework of the Einstein-Cartan model (\ref{LH}).

We begin with the angular momentum conservation and notice that
\begin{eqnarray}
cD{\mathfrak S}_{\alpha\beta} &=& c\widetilde{D}{\mathfrak S}_{\alpha\beta}
+ {\frac 1\kappa}\Big[\overline{a}{}_0 K_\alpha{}^\lambda \wedge T_\lambda\wedge
\vartheta_\beta - \overline{a}{}_0K_\beta{}^\lambda\wedge 
T_\lambda\wedge\vartheta_\alpha\nonumber\\
&& +\,K_\alpha{}^\lambda\wedge\eta_{\lambda\beta\gamma}\wedge T^\gamma
+ K_\beta{}^\lambda\wedge\eta_{\alpha\lambda\gamma}\wedge T^\gamma \Big].\label{DS1}
\end{eqnarray}
We have replaced the spin in the first line by the torsion using the field equation (\ref{ECH2}). One can further simplify the last line in (\ref{DS1}) as follows.

To begin with, we split one term into two halves
\begin{equation}
K_\alpha{}^\lambda\wedge\eta_{\lambda\beta\gamma}\wedge T^\gamma = {\frac 12}K_\alpha{}^\lambda\wedge
\eta_{\lambda\beta\gamma}\wedge T^\gamma + {\frac 12}K_\alpha{}^\lambda\wedge\eta_{\lambda\beta\gamma}
\wedge K^\gamma{}_\nu\wedge\vartheta^\nu.\label{KeT1}
\end{equation}
Next, on the right-hand side we substitute $K_\alpha{}^\lambda = -\,{\frac 12}\eta_\alpha{}^{\lambda\rho\sigma}K^\star_{\rho\sigma}$ into the first term, and $K^\gamma{}_\nu = -\,{\frac 12}\eta^\gamma{}_{\nu\rho\sigma}K^\star{}^{\rho\sigma}$ into the second one. Since $\eta_{\lambda\beta\gamma} = \eta_{\lambda\beta\gamma\delta}\vartheta^\delta$, we evaluate the products of the Levi-Civita tensors in terms of the delta Kroneckers to find
\begin{eqnarray}
K_\alpha{}^\lambda\wedge\eta_{\lambda\beta\gamma}\wedge T^\gamma = {\frac 12}\,g_{\alpha\beta}
\,K^\star_{\rho\sigma}\wedge\vartheta^\sigma\wedge T^\rho\nonumber\\ \label{KeT2}
-\,{\frac 12}(K^\star_{\beta\sigma}\wedge K^\sigma{}_\rho\wedge\vartheta^\rho)\wedge\vartheta_\alpha 
+ {\frac 12}(K_\alpha{}^\sigma\wedge K^\star_{\sigma\rho}\wedge\vartheta^\rho)\wedge\vartheta_\beta.
\end{eqnarray}
Substituting this into the last line of (\ref{DS1}), we derive
\begin{equation}
cD{\mathfrak S}_{\alpha\beta} = c\widetilde{D}{\mathfrak S}_{\alpha\beta} + \vartheta_\alpha\wedge
\Delta {\mathfrak T}_\beta - \vartheta_\beta\wedge\Delta {\mathfrak T}_\alpha,\label{DS2}
\end{equation}
where we introduced, recalling the definition (\ref{Khat}),
\begin{equation}\label{delT}
\Delta {\mathfrak T}^\alpha := {\frac 1{2\kappa}}\left(\widehat{K}^{\alpha\lambda}\wedge 
K_{\lambda\rho} + K^{\alpha\lambda}\wedge \widehat{K}_{\lambda\rho}\right)\wedge\vartheta^\rho.
\end{equation}

Summarizing, we have verified that the conservation law (\ref{Noe2}) of the total angular momentum in the Einstein-Cartan model reads:
\begin{equation}
c\widetilde{D}{\mathfrak S}_{\alpha\beta} + \vartheta_\alpha\wedge\left({\mathfrak T}_\beta +
\Delta {\mathfrak T}_\beta\right) - \vartheta_\beta\wedge\left({\mathfrak T}_\alpha + 
\Delta {\mathfrak T}_\alpha\right) = 0.\label{SconsEC}
\end{equation}

Let us now turn to the discussion of the energy-momentum conservation law. We start with (\ref{conmomC}) which we rewrite as
\begin{eqnarray}\label{conmomC5}
\widetilde{D}{\mathfrak T}_\alpha = {\frac 12}(e_\alpha\rfloor K^{\beta\gamma})\,
c\widetilde{D}{\mathfrak S}_{\beta\gamma} - {\frac 12}(e_\alpha\rfloor \widetilde{D}
K^{\beta\gamma})\!\wedge\! c{\mathfrak S}_{\beta\gamma} +  {\frac 12}(e_\alpha\rfloor 
\widetilde{R}^{\beta\gamma})\!\wedge\! c{\mathfrak S}_{\beta\gamma}.
\end{eqnarray}
Using again the technique of ($\varphi$-$\psi$-$\chi$)-maps (\ref{math}), we recover (\ref{Spsi}) for the spin 3-form, and then for the first term on the right-hand side of (\ref{conmomC5}) we find
\begin{eqnarray}\label{KS1}
{\frac 12}\,(e_\alpha\rfloor K^{\beta\gamma})\,c\widetilde{D}{\mathfrak S}_{\beta\gamma} 
= (e_\alpha\rfloor T^\gamma -K_\alpha{}^\gamma)\wedge\,c\widetilde{D}\{\psi({\mathfrak S})\}_\gamma.
\end{eqnarray}
For the second term on the right-hand side of (\ref{conmomC5}), we make use the field equation (\ref{ECH2}) to derive
\begin{eqnarray}
- {\frac 12}\,(e_\alpha\rfloor \widetilde{D}K^{\beta\gamma})\wedge c{\mathfrak S}_{\beta\gamma} 
&=& {\frac 1\kappa}\left[(e_\alpha\rfloor T^\mu)\!\wedge\!\widetilde{D}
(\widehat{K}_{\mu\nu}\!\wedge\!\vartheta^\nu) + T^\mu\!\wedge\!\widetilde{D}
\widehat{K}_{\mu\alpha}\right].\label{KS2}
\end{eqnarray}
Combining (\ref{KS1}) and (\ref{KS2}) we find for the sum of two terms
\begin{eqnarray}\label{KS3}
{\frac 12}\,(e_\alpha\rfloor K^{\beta\gamma})\,c\widetilde{D}{\mathfrak S}_{\beta\gamma} 
- {\frac 12}\,(e_\alpha\rfloor \widetilde{D}K^{\beta\gamma})\wedge c{\mathfrak S}_{\beta\gamma}
= \widetilde{D}(\Delta {\mathfrak T}_\alpha),
\end{eqnarray}
after some (rather lengthy) algebra. Accordingly, we recast the conservation law (\ref{conmomC5}) of the energy-momentum in the Einstein-Cartan model into 
\begin{eqnarray}\label{TconsEC}
\widetilde{D}\left({\mathfrak T}_\alpha + \Delta{\mathfrak T}_\alpha\right) = {\frac 12}
(e_\alpha\rfloor \widetilde{R}^{\beta\gamma})\!\wedge\! c{\mathfrak S}_{\beta\gamma}.
\end{eqnarray}

\section{Recovering effective energy-momentum current}

One can rearrange the conservation laws of the total angular momentum (\ref{SconsEC}) and of the energy-momentum (\ref{TconsEC}) even further with the help of (\ref{Spsi}). Differentiating the spin current 3-form, we derive
\begin{equation}\label{DSp}
\widetilde{D}{\mathfrak S}_{\alpha\beta} = -\,\vartheta_\alpha\wedge \widetilde{D}\{\psi(
{\mathfrak S})\}_\beta + \vartheta_\beta\wedge  \widetilde{D}\{\psi({\mathfrak S})\}{}_\alpha,
\end{equation}
whereas it is straightforward to compute
\begin{equation}
{\frac 12}(e_\alpha\rfloor \widetilde{R}^{\beta\gamma})\wedge c{\mathfrak S}_{\beta\gamma} =
-\,\widetilde{R}_\alpha{}^\gamma\wedge c\{\psi({\mathfrak S})\}_\gamma = 
c\widetilde{D}\widetilde{D}\{\psi({\mathfrak S})\}_\alpha.\label{RSp}
\end{equation}
Accordingly, we bring the conservation laws (\ref{SconsEC}) and (\ref{TconsEC}) to the form
\begin{eqnarray}
\vartheta_\alpha\!\!\wedge\!\!\left({\mathfrak T}_\beta + \Delta {\mathfrak T}_\beta - 
c\widetilde{D}\{\psi({\mathfrak S})\}_\beta\right) - \vartheta_\beta\!\!\wedge\!\!\left(
{\mathfrak T}_\alpha + \Delta {\mathfrak T}_\alpha - c\widetilde{D}\{\psi(
{\mathfrak S})\}_\alpha\right) = 0,\label{SconsEC1}\\ 
\widetilde{D}\left({\mathfrak T}_\alpha + \Delta{\mathfrak T}_\alpha - c\widetilde{D}
\{\psi({\mathfrak S})\}_\alpha\right) = 0.\label{TconsEC1}
\end{eqnarray}

Making use of (\ref{delT}) and (\ref{Kpsi}), we identify the effective current (\ref{Teff1}) with 
\begin{equation}
{\mathfrak T}{}^{\rm eff}_\alpha = {\mathfrak T}_\alpha + \Delta{\mathfrak T}_\alpha 
- c\widetilde{D}\{\psi({\mathfrak S})\}_\alpha.\label{Teff2}
\end{equation}
Equations (\ref{SconsEC1}) and (\ref{TconsEC1}) show that the effective energy-momentum current is symmetric and conserved:
\begin{equation}\label{TeffC}
\vartheta_\alpha\wedge {\mathfrak T}{}^{\rm eff}_\beta - \vartheta_\beta\wedge 
{\mathfrak T}{}^{\rm eff}_\alpha = 0,\qquad \widetilde{D}{\mathfrak T}{}^{\rm eff}_\alpha = 0.
\end{equation}
This demonstrates the consistency of the effective Einstein theory (\ref{Eeff}).

\subsection{Evaluation of the effective energy-momentum current}

Computation of the effective energy-momentum (\ref{Teff1}) is straightforward but rather lengthy. One needs to find the last term which is a quadratic function of the spin 
\begin{eqnarray}
&{\frac 1{2\kappa}}\left(\widehat{K}_{\alpha\beta}\wedge K^\beta{}_\lambda + K_{\alpha\beta}\wedge 
\widehat{K}^\beta{}_\lambda\right)\wedge\vartheta^\lambda = \zeta\kappa c^2\overline{a}{}_0
\{\chi({\mathfrak S})\}_{\alpha\beta}\wedge\{\chi({\mathfrak S})\}^\beta{}_\lambda\wedge
\vartheta^\lambda&\nonumber\\
&-\,{\frac {\zeta\kappa c^2}{2}}\left[\{\chi({\mathfrak S})\}_{\alpha\beta}\wedge 
\{\chi({\mathfrak S}^\star)\}^\beta{}_\lambda + \{\chi({\mathfrak S}^\star)\}_{\alpha\beta}\wedge 
\{\chi({\mathfrak S})\}^\beta{}_\lambda\right]\wedge\vartheta^\lambda.&\label{Teff3}
\end{eqnarray}

Next, we use (\ref{cP123}) and find for the first term on the right-hand side of (\ref{Teff3}):
\begin{eqnarray}
&\{\chi({\mathfrak S})\}_{\alpha\beta}\wedge\{\chi({\mathfrak S})\}^\beta{}_\lambda\wedge
\vartheta^\lambda = \{\Phi^\star({\mathfrak S})\}_{\alpha\beta}\wedge\{\Phi^\star({\mathfrak S}
)\}^\beta{}_\lambda\wedge\vartheta^\lambda& \nonumber\\
&= {\frac 18}\eta_{\alpha\beta}{}^{\mu\nu}\left({\mathfrak S}_{\mu\nu}{}^\lambda
+ {\mathfrak S}_\mu{}^\lambda{}_\nu + {\mathfrak S}^\lambda{}_{\nu\mu} - 2\delta^\lambda_\mu
{\mathfrak S}_\nu\right)\left({\mathfrak S}_{\gamma\lambda}{}^\beta + {\mathfrak S}_\gamma{}^\beta
{}_\lambda + {\mathfrak S}^\beta{}_{\lambda\gamma}\right)\eta^\gamma .&\label{Teff4}
\end{eqnarray}
The computations above were based on (\ref{Pp}) that yields
\begin{equation}
\{\Phi({\mathfrak S})\}_{\mu\nu}{}^\lambda = {\frac 12}\left({\mathfrak S}_{\mu\nu}{}^\lambda
+ {\mathfrak S}_\mu{}^\lambda{}_\nu + {\mathfrak S}^\lambda{}_{\nu\mu} - \delta^\lambda_\mu
{\mathfrak S}_\nu + \delta^\lambda_\nu {\mathfrak S}_\mu\right),\label{Pp2}
\end{equation}
from which $\{\Phi({\mathfrak S})\}_{\lambda\nu}{}^\lambda = -\,{\frac 12}{\mathfrak S}_\nu$, where the trace of the spin tensor is ${\mathfrak S}_\alpha = {\mathfrak S}_{\mu\alpha}{}^\mu$. 

Now let us analyse the last line in (\ref{Teff3}). We find
\begin{eqnarray}
&-\,\left[\{\chi({\mathfrak S})\}_{\alpha\beta}\wedge\{\chi({\mathfrak S}^\star)\}^\beta{}_\lambda
+ \{\chi({\mathfrak S}^\star)\}_{\alpha\beta}\wedge\{\chi({\mathfrak S})\}^\beta{}_\lambda\right]
\wedge\vartheta^\lambda & \nonumber\\
&= - {\frac 12}\left({\mathfrak S}^{\mu\nu}{}_\alpha + {\mathfrak S}^\mu{}_\alpha{}^\nu 
+ {\mathfrak S}_\alpha{}^{\nu\mu}\right){\mathfrak S}_{\mu\nu}{}^\gamma\eta_\gamma
+ {\frac 14}\left({\mathfrak S}^{\mu\nu}{}_\lambda + {\mathfrak S}^\mu{}_\lambda{}^\nu 
+ {\mathfrak S}_\lambda{}^{\nu\mu}\right){\mathfrak S}_{\mu\nu}{}^\lambda\eta_\alpha&\nonumber\\
&+\,{\mathfrak S}_{\alpha\beta}{}^\gamma{\mathfrak S}^\beta\eta_\gamma - {\frac 12}{\mathfrak S}_\rho
{\mathfrak S}^\rho\eta_\alpha .&\label{Teff7}
\end{eqnarray}
Substituting (\ref{Teff4}) and (\ref{Teff7}) into (\ref{Teff3}), we find the quadratic in spin contribution to the effective energy-momentum current. 

It is straightforward to evaluate the linear in spin contribution:
\begin{eqnarray}
{\frac 1\kappa}\widetilde{D}\left(\widehat{K}_{\alpha\lambda}\wedge\vartheta^\lambda\right)
= - c\widetilde{D}\{\psi({\mathfrak S})\}_\alpha = {\frac c2}\,\widetilde{D}_\nu\left(
{\mathfrak S}^{\mu\nu}{}_\alpha + {\mathfrak S}^\mu{}_\alpha{}^\nu + {\mathfrak S}_\alpha{}^{\nu\mu}
\right)\eta_\mu,\label{Teff8}
\end{eqnarray}
where we used (\ref{Kpsi}) and (\ref{pv}).

Collecting all together, we end up with the effective energy-momentum current (\ref{Teff1}) for any matter sources (${\mathfrak T}_\alpha$, ${\mathfrak S}_{\alpha\beta}$) in extended Einstein-Cartan model (\ref{LH}):
\begin{eqnarray}
{\mathfrak T}^{\rm eff}_\alpha &=&  {\mathfrak T}_\alpha + {\frac c2}\,\widetilde{D}_\nu\left(
{\mathfrak S}^{\mu\nu}{}_\alpha + {\mathfrak S}^\mu{}_\alpha{}^\nu + {\mathfrak S}_\alpha{}^{\nu\mu}
\right)\eta_\mu \nonumber\\
&& +\,{\frac {\zeta\kappa c^2}{8}}\big[ -\,2\left({\mathfrak S}^{\mu\nu}{}_\alpha 
+ {\mathfrak S}^\mu{}_\alpha{}^\nu + {\mathfrak S}_\alpha{}^{\nu\mu}\right){\mathfrak S}_{\mu\nu}
{}^\gamma\eta_\gamma \nonumber\\
&& +\,\left({\mathfrak S}^{\mu\nu}{}_\lambda + {\mathfrak S}^\mu{}_\lambda{}^\nu 
+ {\mathfrak S}_\lambda{}^{\nu\mu}\right){\mathfrak S}_{\mu\nu}{}^\lambda\eta_\alpha 
+\,4{\mathfrak S}_{\alpha\beta}{}^\gamma{\mathfrak S}^\beta\eta_\gamma 
- 2{\mathfrak S}_\rho{\mathfrak S}^\rho\eta_\alpha\big] \nonumber\\
&& +\,{\frac {\zeta\kappa c^2}{8}}\overline{a}{}_0\eta_{\alpha\beta}{}^{\mu\nu}({\mathfrak S}_{\mu\nu}{}^\lambda 
+ {\mathfrak S}_\mu{}^\lambda{}_\nu + {\mathfrak S}^\lambda{}_{\nu\mu} - 2\delta^\lambda_\mu
{\mathfrak S}_\nu)\times \nonumber\\
&& \times \,({\mathfrak S}_{\gamma\lambda}{}^\beta + {\mathfrak S}_\gamma{}^\beta{}_\lambda 
+ {\mathfrak S}^\beta{}_{\lambda\gamma})\eta^\gamma.\label{Teff}
\end{eqnarray}

\subsection{Effective energy-momentum current for spinning fluid}

The energy-momentum current of the Weyssenhoff fluid is described by (\ref{TW}), whereas the spin 3-form (\ref{SW}) satisfies the Frenkel condition $u^\beta{\mathcal S}_{\alpha\beta} = 0$. Hence ${\mathfrak S}_\mu = 0$, and 
\begin{eqnarray}
&-\,{\frac {\zeta\kappa c^2}{2}}\left[\{\chi({\mathfrak S})\}_{\alpha\beta}\wedge 
\{\chi({\mathfrak S}^\star)\}^\beta{}_\lambda + \{\chi({\mathfrak S}^\star)\}_{\alpha\beta}\wedge 
\{\chi({\mathfrak S})\}^\beta{}_\lambda\right]\wedge\vartheta^\lambda&\nonumber\\
&= {\frac {\zeta\kappa}{8}}\,{\mathcal S}_{\mu\nu}{\mathcal S}^{\mu\nu}
\left(- 2u_\alpha u + c^2\,\eta_\alpha\right),&\label{TeW1}\\
& \zeta\kappa c^2\overline{a}{}_0\{\chi({\mathfrak S})\}_{\alpha\beta}\wedge\{\chi(
{\mathfrak S})\}^\beta{}_\lambda\wedge\vartheta^\lambda = {\frac {\zeta\kappa c^2}{4}}
\overline{a}{}_0{\mathcal S}^\star_{\alpha\beta}{\mathcal S}^{\gamma\beta}\eta_\gamma.&\label{TeW2}
\end{eqnarray}
It is easy to verify that ${\mathcal S}^\star_{\alpha\beta}{\mathcal S}^{\gamma\beta} = {\frac 14}\delta_\alpha^\gamma{\mathcal S}^\star_{\mu\nu}{\mathcal S}^{\mu\nu}$, and thus 
\begin{equation}
\Delta{\mathfrak T}_\alpha = -\,{\frac {\zeta\kappa}{4}}{\mathcal S}_{\mu\nu}{\mathcal S}^{\mu\nu}
\,u_\alpha u + {\frac {\zeta\kappa c^2}{8}}\Big({\mathcal S}_{\mu\nu}{\mathcal S}^{\mu\nu} + 
{\frac {\overline{a}{}_0}{2}}{\mathcal S}^\star_{\mu\nu}{\mathcal S}^{\mu\nu}\Big)\eta_\alpha.\label{DTW}
\end{equation}
On the other hand, from (\ref{TW}) we derive
\begin{eqnarray}
{\mathfrak T}_\alpha + {\frac 1\kappa}\widetilde{D}\big(\widehat{K}_{\alpha\lambda}\wedge
\vartheta^\lambda\big) &=& -\,p\big(\eta_\alpha - {\frac 1{c^2}}u_\alpha u\big) 
+ {\frac {\varepsilon}{c^2}}\,u_\alpha u + {\frac 1{c^2}}u^\lambda\widetilde{D}_\nu(
{\mathcal S}_{\alpha\lambda}u^\nu)\,u^\mu\eta_\mu\nonumber\\
&& + {\frac 12}\,\widetilde{D}_\nu\left({\mathcal S}^{\mu\nu}u_\alpha 
+ {\mathcal S}^\mu{}_\alpha u^\nu + {\mathcal S}_\alpha{}^\nu u^\mu\right)\eta_\mu .\label{STW}
\end{eqnarray}
The sum of (\ref{DTW}) and (\ref{STW}) yields the effective energy-momentum current. 

Making use of the equation of motion of spin (\ref{spineq}) which can be written as
\begin{equation}
\widetilde{D}_\nu({\mathcal S}_{\alpha\beta}u^\nu) - {\frac 1{c^2}}u_\beta u^\gamma \widetilde{D}_\nu
({\mathcal S}_{\alpha\gamma}u^\nu) - {\frac 1{c^2}}u_\alpha u^\gamma \widetilde{D}_\nu
({\mathcal S}_{\gamma\beta}u^\nu) = 0,\label{spineqEC}
\end{equation}  
we can rearrange the terms with derivatives of spin in (\ref{STW}). As a result, the effective energy-momentum current for the Weyssenhoff fluid reads
\begin{eqnarray}
{\mathfrak T}^{\rm eff}_\alpha = -\,p^{\rm eff}\big(\eta_\alpha - {\frac 1{c^2}}u_\alpha u\big) 
+ {\frac {\varepsilon^{\rm eff}}{c^2}}\,u_\alpha u + 
\Big(g^{\nu\lambda} + {\frac 1{c^2}}u^\nu u^\lambda\Big)\widetilde{D}_\nu\left(
u_{(\mu}{\mathcal S}_{\alpha)\lambda}\right)\eta^\mu,\label{TeWF}
\end{eqnarray}
where the effective pressure and energy density depend on spin:
\begin{eqnarray}
p^{\rm eff} &=& p -  {\frac {\zeta\kappa c^2}{8}}\Big({\mathcal S}_{\mu\nu}{\mathcal S}^{\mu\nu}
+ {\frac {\overline{a}{}_0}{2}}{\mathcal S}^\star_{\mu\nu}{\mathcal S}^{\mu\nu}\Big),\label{Peff}\\
\varepsilon^{\rm eff} &=& \varepsilon -  {\frac {\zeta\kappa c^2}{8}}\Big(
{\mathcal S}_{\mu\nu}{\mathcal S}^{\mu\nu} - 
{\frac {\overline{a}{}_0}{2}}{\mathcal S}^\star_{\mu\nu}{\mathcal S}^{\mu\nu}\Big).\label{Epeff}
\end{eqnarray}
Although formally these quantities contain ${\mathcal S}^\star_{\mu\nu}{\mathcal S}^{\mu\nu}$ invariant, the latter vanishes in view of the Frenkel condition $u_\mu{\mathcal S}^{\mu\nu} = 0$. Accordingly, the parity-odd coupling constant $\overline{a}{}_0$ appears in the effective Einstein field equations (\ref{Eeff}) only via the parameter $\zeta$, given by (\ref{zeta0}), which thus determines the ``strength'' of spin-spin contributions, (\ref{Peff}) and (\ref{Epeff}), in the effective energy-momentum current of the fluid (\ref{TeWF}).

\section{Torsion-square Poincar\'e gravity model}

Let us consider the generalization of the Einstein-Cartan model with the Lagrangian that contains all possible quadratic invariants of the torsion:
\begin{equation}
V = {\frac {1}{2\kappa c}}\Bigl\{\left(\eta_{\alpha\beta} + \overline{a}{}_0\vartheta_\alpha
\wedge\vartheta_\beta\right)\wedge R^{\alpha\beta} - 2\lambda_0\eta 
- \,T^\alpha\wedge\sum_{I=1}^3\left[a_I\,{}^*({}^{(I)}T_\alpha)
+ \overline{a}_I\,{}^{(I)}T_\alpha\right]\Bigr\}.\label{LQT}
\end{equation}
For completeness, we included the cosmological constant $\lambda_0$ term. As compared to (\ref{LH}), the new Lagrangian contains 5 additional (dimensionless) coupling constants: $a_1, a_2, a_3$ and $\overline{a}_1, \overline{a}_2 = \overline{a}_3$. The two latter parity-odd constants are equal because the two last terms in (\ref{LQT}) are the same: 
\begin{equation}\label{T23}
T^\alpha\wedge{}^{(2)}T_\alpha = T^\alpha\wedge{}^{(3)}T_\alpha = {}^{(2)}T^\alpha\wedge{}^{(3)}T_\alpha,
\end{equation}
whereas $T^\alpha\wedge{}^{(1)}T_\alpha = {}^{(1)}T^\alpha\wedge{}^{(1)}T_\alpha$. One can prove these relations directly from the definitions (\ref{iT23})-(\ref{iT1}).

The gauge gravitational field equations are derived from the variation of the total Lagrangian $V + {\frac 1c}L$ with respect to the coframe and the local Lorentz connection:
\begin{eqnarray}
{\frac 12}\eta_{\alpha\beta\gamma}\wedge R^{\beta\gamma} + \overline{a}{}_0R_{\alpha\beta}
\wedge\vartheta^\beta - \lambda_0\eta_\alpha - Dh_\alpha + q^{(T)}_\alpha =
\kappa\,{\mathfrak T}_\alpha,\label{EQT1}\\
\eta_{\alpha\beta\gamma}\wedge T^{\gamma} + \overline{a}{}_0\left(T_\alpha\wedge\vartheta_\beta
- T_\beta\wedge\vartheta_\alpha \right) + h_\alpha\wedge\vartheta_\beta
- h_\beta\wedge\vartheta_\alpha = \kappa c\,{\mathfrak S}_{\alpha\beta}.\label{EQT2}
\end{eqnarray}
Here we denoted the linear and quadratic functions of the torsion
\begin{equation}
h_\alpha = \sum_{I=1}^3\left[a_I\,{}^*({}^{(I)}T_\alpha) 
+ \overline{a}_I\,{}^{(I)}T_\alpha\right],\quad 
q^{(T)}_\alpha = {\frac 12}\left[(e_\alpha\rfloor T^\beta)\wedge h_\beta - T^\beta\wedge 
e_\alpha\rfloor h_\beta\right].\label{qa}
\end{equation}
It is straightforward to prove the simple properties of these objects which follow directly from their definitions:
\begin{equation}
\vartheta^\alpha\wedge q^{(T)}_\alpha = 0,\qquad 
\vartheta^\alpha\wedge h_\alpha = -\,a_2{}^*T + \overline{a}_3{}^*\overline{T},
\qquad e^\alpha\rfloor h_\alpha = a_3 \overline{T} + \overline{a}_2 T.\label{eth}
\end{equation}
It is important to notice that the forms (\ref{qa}) satisfy the geometrical identity 
\begin{equation}
h_\alpha\wedge T_\beta - h_\beta\wedge T_\alpha + q^{(T)}_\alpha\wedge\vartheta_\beta - 
q^{(T)}_\beta\wedge\vartheta_\alpha \equiv 0.\label{hTq}
\end{equation}
To verify this, we notice that $h_\alpha$ is a linear combination of the irreducible parts of the torsion and its dual. The relation (\ref{hTq}) is valid always irrespectively whether the field equations are fulfilled or not. 

As we see, the Einstein-Cartan field equations (\ref{ECH1}) and (\ref{ECH2}) are now replaced by the system (\ref{EQT1}) and (\ref{EQT2}) modified by the presence of the many additional torsion-dependent terms. However, the generalized quadratic Poincar\'e gravity model (\ref{EQT1})-(\ref{EQT2}) is not different dynamically from the Einstein-Cartan theory. 

In particular, the relation between the spin and the torsion is still algebraic one, and we can solve (\ref{EQT2}) for the torsion as a function of spin. Using (\ref{iT23})-(\ref{iT1}), we recast (\ref{EQT2}) into
\begin{equation}
{}^*\Bigl(-\,{}^{(1)}T^\alpha + 2{}^{(2)}T^\alpha + {\frac 12}{}^{(3)}T^\alpha\Bigr)
-\,\overline{a}{}_0\Big({}^{(1)}T^\alpha + {}^{(2)}T^\alpha + {}^{(3)}T^\alpha\Big) 
= \kappa c\left\{\psi({\mathfrak S})\right\}^\alpha + h^\alpha.\label{ECHQT}
\end{equation}
Let us solve this equation in terms of the torsion. The latter is a sum of the three irreducible pieces, and to find them we insert (\ref{qa}) into (\ref{ECHQT}). The result reads
\begin{eqnarray}
-\,(1 + a_1){}^*({}^{(1)}T^\alpha) + (2 - a_2){}^*({}^{(2)}T^\alpha) + \Big({\frac 12} - a_3\Big)
{}^*({}^{(3)}T^\alpha)&&\nonumber\\
- \Big(\overline{a}{}_0 +\overline{a}_1\Big){}^{(1)}T^\alpha - \Big(\overline{a}{}_0 + \overline{a}_2\Big)
{}^{(2)}T^\alpha - \Big(\overline{a}{}_0 + \overline{a}_3\Big){}^{(3)}T^\alpha 
&=& \kappa c\left\{\psi({\mathfrak S})\right\}^\alpha.\label{ECQT}
\end{eqnarray}
Taking the Hodge dual and combining the result with (\ref{ECQT}), we find the irreducible torsion parts (recall the equality $\overline{a}_2 = \overline{a}_3$):
\begin{eqnarray}
{}^{(1)}T^\alpha &=& {\frac {\kappa c}{(1 + a_1)^2 + (\overline{a}{}_0 + \overline{a}_1)^2}}
\left\{(1 + a_1) {}^{(1)}\widehat{\mathfrak S}^\alpha + (\overline{a}{}_0 +\overline{a}_1)
{}^*({}^{(1)}\widehat{\mathfrak S}^\alpha)\right\},\label{QTsol1}\\
{}^{(2)}T^\alpha &=& {\frac {-\,\kappa c}{(2 - a_2)(1 - 2a_3) + 2(\overline{a}{}_0
+ \overline{a}_2)^2}} \left\{(1 - 2a_3){}^{(2)}\widehat{\mathfrak S}^\alpha 
+ (\overline{a}{}_0 +\overline{a}_2){}^*({}^{(3)}\widehat{\mathfrak S}^\alpha)
\right\},\label{QTsol2}\\
{}^{(3)}T^\alpha &=& {\frac {\kappa c}{(2 - a_2)(1 - 2a_3) + 2(\overline{a}{}_0 + \overline{a}_2)^2}}
\left\{(2 - a_2)\,{}^{(3)}\widehat{\mathfrak S}^\alpha + 2(\overline{a}{}_0
+\overline{a}_2){}^*({}^{(2)}\widehat{\mathfrak S}^\alpha)\right\}. \label{QTsol3}
\end{eqnarray}
Here from the components of the (tensor-valued) 3-form ${\mathfrak S}_{\alpha\beta} = {\mathfrak S}_{\alpha\beta}{}^\mu\eta_\mu$ of the spin current we construct the (vector-valued) 2-form $\widehat{\mathfrak S}^\mu := {\frac 12}{\mathfrak S}_{\alpha\beta}{}^\mu\,\vartheta^\alpha\wedge\vartheta^\beta$, and decompose the latter into the three irreducible parts, using the decomposition of the torsion 2-form (\ref{iT23})-(\ref{iT1}) as a pattern.

\section{Effective Einstein's equation in torsion-square model}

After we have solved Cartan's equation (\ref{EQT2}), we can substitute the torsion as a function of spin into (\ref{EQT1}) and recast the latter into an effective Einstein's equation. The decompositions (\ref{etaR2}) and (\ref{Rth}) are valid in general, and we can use them to rewrite (\ref{EQT1}) as
\begin{equation}
{\frac 12}\eta_{\alpha\beta\gamma}\wedge \widetilde{R}^{\beta\gamma} - \lambda_0\eta_\alpha
= \kappa\,{\mathfrak T}^{\rm eff}_\alpha.\label{ETQeff}
\end{equation}
The effective energy-momentum current 3-form now reads
\begin{eqnarray}
{\mathfrak T}^{\rm eff}_\alpha = {\mathfrak T}_\alpha + {\frac 1\kappa}\widetilde{D}
\left(\widehat{K}_{\alpha\lambda}\wedge\vartheta^\lambda + h_\alpha\right) 
+ \,\Delta{\mathfrak T}_\alpha + {\frac 1\kappa}\left(K_\alpha{}^\beta h_\beta 
- q^{(T)}_\alpha\right),\label{TQTeff}
\end{eqnarray}
where $\Delta{\mathfrak T}_\alpha$ was introduced in (\ref{delT}). This current contains the terms linear in the torsion (under the derivative) and the torsion-square terms (the last three). 

Let us consider the linear terms. We have ${\frac 12}K^{\mu\nu}\wedge\eta_{\alpha\mu\nu} = {\frac 12}K^{\mu\nu}\eta_{\alpha\mu\nu\beta}\wedge\vartheta^\beta = K^\star_{\alpha\beta}\wedge\vartheta^\beta$, and thus we recast (\ref{ECHQT}) into
\begin{equation}\label{KstSh}
-\,\widehat{K}_{\alpha\beta}\wedge\vartheta^\beta = \kappa c\left\{\psi({\mathfrak S})\right\}^\alpha 
+ h^\alpha,
\end{equation}
and consequently, with the help of (\ref{pv}) we derive
\begin{eqnarray}
{\frac 1\kappa}\widetilde{D}\left(\widehat{K}_{\alpha\lambda}\wedge\vartheta^\lambda + h_\alpha\right)
= -\,c\widetilde{D}\{\psi({\mathfrak S})\}_\alpha 
= {\frac c2}\,\widetilde{D}_\nu\left({\mathfrak S}^{\mu\nu}{}_\alpha 
+ {\mathfrak S}^\mu{}_\alpha{}^\nu + {\mathfrak S}_\alpha{}^{\nu\mu}\right)\eta_\mu.\label{DQTeff}
\end{eqnarray}

The effective energy-momentum current is symmetric and covariantly conserved, (\ref{TeffC}). To demonstrate the angular momentum conservation, we notice that
\begin{eqnarray}
cD{\mathfrak S}_{\alpha\beta} 
&=& c\widetilde{D}{\mathfrak S}_{\alpha\beta} + {\frac 1\kappa}\Big[\overline{a}{}_0
K_\alpha{}^\lambda\wedge T_\lambda\wedge\vartheta_\beta - \overline{a}{}_0K_\beta{}^\lambda\wedge 
T_\lambda\wedge\vartheta_\alpha\nonumber\\
&& +\,K_\alpha{}^\lambda\wedge\eta_{\lambda\beta\gamma}\wedge T^\gamma + K_\beta{}^\lambda\wedge 
\eta_{\alpha\lambda\gamma}\wedge T^\gamma + K_\alpha{}^\lambda\wedge h_\lambda\wedge
\vartheta_\beta \nonumber\\
&& -\,K_\alpha{}^\lambda\wedge h_\beta\wedge\vartheta_\lambda + K_\beta{}^\lambda\wedge h_\alpha
\wedge\vartheta_\lambda - K_\beta{}^\lambda\wedge h_\lambda\wedge\vartheta_\alpha\Big].\label{DSQ1}
\end{eqnarray}
We have replaced the spin in the first line by the torsion using the field equation (\ref{EQT2}). One can further simplify the right-hand side of (\ref{DSQ1}) by noticing that $K^\alpha{}_\lambda\wedge\vartheta^\lambda = T^\alpha$, and making use of the derivations (\ref{KeT1})-(\ref{delT}) and the identity (\ref{hTq}), to find 
\begin{equation}
c\widetilde{D}{\mathfrak S}_{\alpha\beta} + \vartheta_\alpha\wedge\Big[\Delta {\mathfrak T}_\beta
+ {\frac 1\kappa}(K_\beta{}^\lambda\wedge h_\lambda - q^{(T)}_\beta)\Big] 
- \vartheta_\beta\wedge\Big[\Delta {\mathfrak T}_\alpha + {\frac 1\kappa}
(K_\alpha{}^\lambda\wedge h_\lambda - q^{(T)}_\alpha)\Big] = 0.\label{SconsQ}
\end{equation}
Then, with the help of (\ref{Spsi}), (\ref{DSp}) and (\ref{KstSh}), we recast the angular momentum conservation (\ref{SconsEC}) into $\vartheta_\alpha\wedge {\mathfrak T}{}^{\rm eff}_\beta - \vartheta_\beta\wedge {\mathfrak T}{}^{\rm eff}_\alpha = 0$ for the effective energy-momentum current (\ref{TQTeff}).

In a similar way, we can generalize the derivations (\ref{conmomC5})-(\ref{TconsEC}) to demonstrate that the effective energy-momentum current is conserved, $\widetilde{D}{\mathfrak T}{}^{\rm eff}_\alpha = 0$, thereby verifying completely (\ref{TeffC}) for the torsion-square Poincar\'e gravity model. 

Substituting the torsion (\ref{QTsol1})-(\ref{QTsol3}) into (\ref{TQTeff}) we obtain a generalization of (\ref{Teff}). The resulting expression is very bulky in general case of an arbitrary matter source, and we do not give it here. For the physically interesting special cases of the spinning fluid and the Dirac fermion field, the resulting effective theories are as follows.

\subsection{Effective fluid dynamics in torsion-square model}

For the Weyssenhoff spinning fluid, all the three irreducible torsion parts are nontrivial. However, computations are greatly simplified due to the Frenkel condition $u^\beta{\mathcal S}_{\alpha\beta} = 0$. The effective energy-momentum current for the spinning fluid reads
\begin{eqnarray}
{\mathfrak T}^{\rm eff}_\alpha = -\,p^{\rm eff}\big(\eta_\alpha - {\frac 1{c^2}}u_\alpha u\big) 
+ {\frac {\varepsilon^{\rm eff}}{c^2}}\,u_\alpha u + 
\Big(g^{\nu\lambda} + {\frac 1{c^2}}u^\nu u^\lambda\Big)\widetilde{D}_\nu\left(
u_{(\mu}{\mathcal S}_{\alpha)\lambda}\right)\eta^\mu,\label{QTeWF}
\end{eqnarray}
where the effective pressure and energy density depend on spin:
\begin{eqnarray}
p^{\rm eff} =p - {\frac {\zeta\kappa c^2}{8}}\,{\mathcal S}_{\mu\nu}{\mathcal S}^{\mu\nu},\qquad
\varepsilon^{\rm eff} = \varepsilon -  {\frac {\zeta\kappa c^2}{8}}\,{\mathcal S}_{\mu\nu}
{\mathcal S}^{\mu\nu}.\label{QEpeff}
\end{eqnarray}
Here we denoted a combination of the coupling constants:
\begin{eqnarray}
\zeta = {\frac {4(1 + a_1)}{3(1 + a_1)^2 + 3(\overline{a}{}_0 + \overline{a}_1)^2}} 
- {\frac {2 - a_2}{3(2 - a_2)(1 - 2a_3) + 6(\overline{a}{}_0 + \overline{a}_2)^2}}.\label{zetaT}
\end{eqnarray}
When the torsion-square terms are absent, $a_1 = a_2 = a_3 = \overline{a}_1 = \overline{a}_2 =0$, we recover the value (\ref{zeta0}) of the Einstein-Cartan theory. Note that there exists a large class of models with the torsion quadratic Lagrangians which yield $\zeta = 0$.

As for spin dynamics, substituting the torsion (\ref{QTsol1})-(\ref{QTsol3}) into (\ref{spineq}), we find (\ref{spineqEC}).

\subsection{Effective Dirac fermion dynamics in torsion-square model}

For the Dirac fermion source with spin (\ref{Tau}), the torsion (\ref{QTsol1})-(\ref{QTsol3}) solution now reads: ${}^{(1)}T^\alpha = 0$, and 
\begin{eqnarray}
{}^{(2)}T^\alpha &=& -\,{\frac {\overline{\zeta}\kappa c\hbar}2}\,
\vartheta^\alpha\wedge\overline{\Psi}\gamma\gamma_5\Psi,\label{QSTsol2}\\
{}^{(3)}T^\alpha &=& -\,{\frac {\zeta\kappa c\hbar}2}\,{}^*(\vartheta^\alpha\wedge\overline{\Psi}
\gamma\gamma_5\Psi),\label{QSTsol3}
\end{eqnarray}
where we denoted
\begin{eqnarray}
\overline{\zeta} &=& {\frac {\overline{a}{}_0 +\overline{a}_2}{(2 - a_2)(1 - 2a_3)
+ 2(\overline{a}{}_0 + \overline{a}_2)^2}},\label{ozetaF}\\
\zeta &=& {\frac {2 - a_2}{(2 - a_2)(1 - 2a_3) + 2(\overline{a}{}_0
+ \overline{a}_2)^2}}.\label{zetaF}
\end{eqnarray}
Although the Dirac spin is totally antisymmetric (\ref{Tau}), in a general torsion-square model (\ref{LQT}) the spacetime torsion has not only the axial trace part, but also the vector trace part is nontrivial. The latter is generated via (\ref{ozetaF}), when the parity-odd terms are present in the Lagrangian. Only for the purely parity-even Poincar\'e model with $\overline{a}{}_0 = 0$ and $\overline{a}_I = 0$, the second irreducible part (\ref{QSTsol2}) vanishes, and the torsion becomes completely antisymmetric. 

Nevertheless, despite the presence of the torsion trace part, Dirac fermions do not feel it, since its spin is coupled solely to the axial torsion part (\ref{QSTsol3}). As a result, the effective dynamics of the Dirac field in the torsion-square model (\ref{LQT}) is determined by the parameter (\ref{zetaF}) which enters the right-hand side of (\ref{QSTsol3}). Substituting (\ref{QSTsol2}) and (\ref{QSTsol3}) into (\ref{TQTeff}) we obtain the effective energy-momentum current
\begin{eqnarray}
{\mathfrak T}_\alpha^{\rm eff} &=& {\frac {i\hbar c}{4}}\left\{\overline{\Psi}\,{}^\ast\!\gamma 
\widetilde{D}_\alpha\Psi + \overline{\Psi}\gamma_\alpha\,{}^\ast\!\widetilde{D}\Psi - \widetilde{D}_\alpha\overline{\Psi}
\,{}^\ast\!\gamma\Psi - {}^\ast\!(\widetilde{D}\overline{\Psi})\,\gamma_\alpha\Psi\right\}\nonumber\\
&& +\,{\frac 3{16}}\zeta\kappa c^2\hbar^2\,e_\alpha\rfloor\left\{(\overline{\Psi}\gamma
\gamma_5\Psi)\wedge{}^*(\overline{\Psi}\gamma\gamma_5\Psi)\right\}.\label{TFeff}
\end{eqnarray}
Finally, inserting the torsion (\ref{QSTsol2})-(\ref{QSTsol3}) into the Dirac equation (\ref{dirRCa}), we recast the latter into an effective nonlinear spinor equation  
\begin{eqnarray}
\hbar{}^\ast\gamma\wedge\left\{i\widetilde{D}\,\Psi + \hbox{$\scriptstyle\frac{3}{8}$}
\zeta\kappa c \hbar\,(\overline{\Psi}\gamma\gamma_5\Psi)\,\gamma_5\Psi\right\} 
+ {}^\ast mc\,\Psi = 0.\label{direff}
\end{eqnarray}
It is worthwhile to note that when the torsion-square terms are absent, $a_I = \overline{a}_I = 0$. the parameter (\ref{zetaF}) reduces to the value (\ref{zeta0}) of the Einstein-Cartan model.

\section{Conclusion}

In this paper we study the parity violation issue in the Poincar\'e gauge theory of gravity. The two classes of models are considered: the extended Einstein-Cartan theory with the so-called Holst term (\ref{LH}) and the torsion-square PG model with the most general Lagrangian (\ref{LQT}) which is constructed of all possible quadratic torsion invariants, including the complete parity-odd sector.

We focus on the analysis of the conservation laws of the corresponding matter sources of the gauge gravitational field: the conservation of the energy-momentum current and the conservation of the total angular momentum. In both PG models, the canonical spin current is coupled algebraically to the spacetime torsion, and by solving the second field equation, we find the torsion as a function of the spin of matter in terms of its irreducible parts (\ref{QTsol1})-(\ref{QTsol3}). After deriving the effective Einstein field equation, we study the structure and the properties of the effective energy-momentum current for arbitrary matter sources. 

By specializing to the macroscopic matter (the Weyssenhoff spinning fluid) and to the microscopic matter (the Dirac fermion field), we demonstrate how the parity violating coupling constants contribute to the strength of an effective spin-spin interaction. One can apply the results obtained for the study of the early and late stages of the cosmological evolution, generalizing the earlier findings \cite{Trautman:1973,Kerlick,Pop2}.

\section*{Acknowledgments}

This work was partially supported by the Russian Foundation for Basic Research (Grant No. 18-02-40056-mega). I am grateful to Friedrich Hehl and Dirk Puetzfeld for the most useful comments and advice.

\appendix

\section{Geometrical definitions}

The gravitational field is described by the coframe $\vartheta^\alpha = e^\alpha_i dx^a$ and connection $\Gamma^{\alpha\beta} = \Gamma_i{}^{\alpha\beta} dx^i$ 1-forms. The translational and rotational field strengths read 
\begin{eqnarray}
T^\alpha &=& D\vartheta^\alpha = d\vartheta^\alpha +\Gamma_\beta{}^\alpha\wedge
\vartheta^\beta,\label{Tor}\\ \label{Cur}
R^{\alpha\beta} &=& d\Gamma^{\alpha\beta} + \Gamma_\gamma{}^\beta\wedge\Gamma^{\alpha\gamma}.
\end{eqnarray}
In components, $T^\alpha = {\frac 12}T_{ij}{}^\alpha dx^i\wedge dx^j$ and 
$R^{\alpha\beta} = {\frac 12}R_{ij}{}^{\alpha\beta} dx^i\wedge dx^j$. The Ricci 1-form:
\begin{equation}
{\rm Ric}^\alpha = e_\beta\rfloor R^{\alpha\beta} = {\rm Ric}_i{}^\alpha dx^i,\qquad
{\rm Ric}_i{}^\alpha = e^j_\beta R_{ji}{}^{\alpha\beta}.\label{Ric}
\end{equation}
The curvature scalar is defined by $R = e_\alpha\rfloor{\rm Ric}^\alpha = e_\alpha\rfloor e_\beta\rfloor R^{\alpha\beta} = e^i_\beta e^j_\alpha R_{ij}{}^{\alpha\beta}$. The Riemannian connection 1-form $\widetilde{\Gamma}_\beta{}^\alpha$ is uniquely defined from the vanishing torsion condition $d\vartheta^\alpha + \widetilde{\Gamma}_\beta{}^\alpha\wedge\vartheta^\beta = 0$. One can decompose the Riemann-Cartan connection 
\begin{equation}
\Gamma^{\alpha\beta} = \widetilde{\Gamma}^{\alpha\beta} - K^{\alpha\beta}\label{GG}
\end{equation}
into the Riemannian and the post-Riemannian parts. The contortion 1-form $K^{\alpha\beta}
= - \,K^{\beta\alpha}$ is algebraically related to the torsion:
\begin{equation}
T^\alpha = K^\alpha{}_\beta\wedge\vartheta^\beta.\label{TK}
\end{equation}
Explicitly, we have for the contortion 1-form:
\begin{equation}\label{Kab}
K_{\alpha\beta} = {\frac 12}\left(e_\alpha\rfloor T_\beta - e_\beta\rfloor T_\alpha
- \vartheta^\gamma\,e_\alpha\rfloor e_\beta\rfloor T_\gamma\right).
\end{equation}
Substituting (\ref{GG}) into (\ref{Cur}), we decompose the curvature 2-form into the the Riemannian and the post-Riemannian parts:
\begin{equation}
R^{\alpha\beta} = \widetilde{R}^{\alpha\beta} - \widetilde{D}K^{\alpha\beta} + 
K_\gamma{}^\beta\wedge K^{\alpha\gamma}.\label{RR}
\end{equation}
Hereafter the Riemannian objects and operators (constructed with the help of the Riemannian connection) are denoted by the tilde. 

Directly from the definitions (\ref{Tor}) and (\ref{Cur}) we derive the Bianchi identities:
\begin{eqnarray}
DR^{\alpha\beta} = 0,\qquad DT^\alpha = R_\beta{}^\alpha\wedge\vartheta^\beta.\label{Bianchi2}
\end{eqnarray}

\section{Torsion decomposition}

The torsion 2-form can be decomposed into the three irreducible pieces, $T^{\alpha}={}^{(1)}T^{\alpha} + {}^{(2)}T^{\alpha} + {}^{(3)}T^{\alpha}$, where
\begin{eqnarray}
{}^{(2)}T^{\alpha} &=& {\frac 13}\vartheta^{\alpha}\wedge T,\qquad 
{}^{(3)}T^{\alpha} = {\frac 13}e^\alpha\rfloor{}^\ast \overline{T},\label{iT23}\\
{}^{(1)}T^{\alpha} &=& T^{\alpha}-{}^{(2)}T^{\alpha} - {}^{(3)}T^{\alpha}.\label{iT1}
\end{eqnarray}
Here the 1-forms of the torsion trace and axial trace are introduced:
\begin{equation}
T := e_\nu\rfloor T^\nu,\qquad \overline{T} := {}^*(T^{\nu}\wedge\vartheta_{\nu}).\label{traces1}
\end{equation}

For the irreducible pieces of the dual torsion ${}^*T^{\alpha} = {}^{(1)}({}^*T^{\alpha}) + {}^{(2)}({}^*T^{\alpha}) + {}^{(3)}({}^*T^{\alpha})$, we have the properties 
\begin{equation}
{}^{(1)}({}^*T^\alpha)={}^*({}^{(1)}T^\alpha),\quad
{}^{(2)}({}^*T^\alpha)={}^*({}^{(3)}T^\alpha),\quad
{}^{(3)}({}^*T^\alpha)={}^*({}^{(2)}T^\alpha).\label{dTdual}
\end{equation}

\section{Algebraic maps: ($\varphi$-$\psi$-$\chi$) technique}\label{math}

In four dimensions, skew-symmetric tensor-valued 3-forms ${\stackrel {(3)}\varphi}_{\alpha\beta} = -\,{\stackrel {(3)}\varphi}_{\beta\alpha}$, vector-valued 2-forms ${\stackrel {(2)}\psi}_\alpha$, and skew-symmetric tensor-valued 1-forms ${\stackrel {(1)}\chi}_{\alpha\beta} = -\,{\stackrel {(1)}\chi}_{\beta\alpha}$ have the same number of independent components. As a result, it is possible to establish one-to-one relations between the spaces of such objects:
\begin{equation}
{\stackrel {(3)}\varphi}_{\alpha\beta}\longleftrightarrow{\stackrel {(2)}\psi}_\alpha
\longleftrightarrow{\stackrel {(1)}\chi}_{\alpha\beta}.\label{maps}
\end{equation}
We define these maps by the linear equations
\begin{eqnarray}\label{map23}
\varphi^{\alpha\beta} &=& \vartheta^\alpha\wedge\psi^\beta - \vartheta^\beta\wedge\psi^\alpha,\\
\psi^\alpha &=& \chi^\alpha{}_\beta\wedge\vartheta^\beta.\label{map12}
\end{eqnarray}
The inverse maps to (\ref{map23}) and (\ref{map12}) are easily found:
\begin{eqnarray}\label{map32}
\psi_\alpha &=& -\,e^\beta\rfloor\varphi_{\alpha\beta} + {\frac 14} \vartheta_\alpha\wedge
e^\beta\rfloor e^\gamma\rfloor\varphi_{\beta\gamma},\\
\chi^\alpha{}_\beta &=& {\frac 12}\left(e^\alpha\rfloor\psi_\beta - e_\beta\rfloor
\psi^\alpha - \vartheta^\gamma e^\alpha\rfloor e_\beta\rfloor\psi_\gamma\right).\label{map21}
\end{eqnarray}
With the help of the relations (\ref{map23}), (\ref{map12}), (\ref{map32}) and (\ref{map21}) we then find the maps between the skew-symmetric tensor-valued 3- and 1-forms:
\begin{eqnarray}\label{map31}
\varphi^{\alpha\beta} &=& \chi^\alpha{}_\gamma\wedge\vartheta^\beta\wedge\vartheta^\gamma 
- \chi^\beta{}_\gamma\wedge\vartheta^\alpha\wedge\vartheta^\gamma,\\
\chi^\alpha{}_\beta &=& {\frac 12}\Big(e^\alpha\rfloor e^\gamma\rfloor\varphi_{\gamma\beta}
- e_\beta\rfloor e_\gamma\rfloor\varphi^{\gamma\alpha} - \vartheta^\delta e^\alpha\rfloor 
e_\beta\rfloor e^\gamma\rfloor\varphi_{\gamma\delta}\nonumber\\
&& +\,{\frac 12}\vartheta^\alpha e_\beta\rfloor e^\gamma\rfloor e^\delta\rfloor
\varphi_{\gamma\delta} - {\frac 12}\vartheta_\beta e^\alpha\rfloor e^\gamma\rfloor 
e^\delta\rfloor\varphi_{\gamma\delta}\Big).\label{map13}
\end{eqnarray}

Technically, we will take a 3-form $\varphi_{\alpha\beta}$ as the starting point, and then 
denote the resulting 2- and 1-forms, obtained from the maps (\ref{map32}) and 
(\ref{map13}), by 
\begin{equation}\label{pcv}
\left\{\psi(\varphi)\right\}_\alpha,\qquad \left\{\chi(\varphi)\right\}^\alpha{}_\beta.
\end{equation}
In components, these maps read as follows. Decomposing $\varphi_{\alpha\beta} = \varphi_{\alpha\beta}{}^\mu\eta_\mu$, and denoting the trace $\varphi_\alpha := \varphi_{\mu\alpha}{}^\mu$, we have explicitly
\begin{eqnarray}
\left\{\psi(\varphi)\right\}_\alpha &=& {\frac 14}\left(\varphi_{\alpha\mu\nu} 
+ \varphi_{\nu\alpha\mu} + \varphi_{\nu\mu\alpha}\right)\eta^{\mu\nu},\label{pv}\\
\left\{\chi(\varphi)\right\}^\alpha{}_\beta &=& {\frac 14}\eta^{\mu\nu\alpha}{}_\beta\left(
\varphi_{\lambda\mu\nu} + \varphi_{\nu\lambda\mu} + \varphi_{\nu\mu\lambda} - g_{\lambda\nu}
\varphi_\mu + g_{\lambda\mu}\varphi_\nu\right)\vartheta^\lambda,\label{cv}
\end{eqnarray}

Technically, it is convenient to introduce an auxiliary tensor-valued 1-form
\begin{equation}
\left\{\Phi(\varphi)\right\}_{\alpha\beta} := {\frac 12}\left(\varphi_{\alpha\beta\lambda}
+ \varphi_{\alpha\lambda\beta} + \varphi_{\lambda\beta\alpha} - g_{\lambda\alpha}\varphi_\beta 
+ g_{\lambda\beta}\varphi_\alpha\right)\vartheta^\lambda.\label{Pp}
\end{equation}
This is not a new independent object, but using it one simplify computations. In particular, we can verify that (recall the right dual $\varphi^\star_{\alpha\beta} = {\frac 12}\eta_{\alpha\beta\rho\sigma}\varphi^{\rho\sigma}$)
\begin{equation}
\left\{\chi(\varphi)\right\}_{\alpha\beta} = -\,\left\{\Phi^\star(\varphi)
\right\}_{\alpha\beta},\ 
\left\{\chi^\star(\varphi)\right\}_{\alpha\beta} = \left\{\Phi(\varphi)
\right\}_{\alpha\beta},\ 
\left\{\chi(\varphi^\star)\right\}_{\alpha\beta} = \left\{\Phi(\varphi)
\right\}_{\alpha\beta}.\label{cP123}
\end{equation}
Consequently, we find the following identities 
\begin{eqnarray}
\left\{\chi^\star(\varphi)\right\}_{\alpha\beta} = \left\{\chi(\varphi^\star)
\right\}_{\alpha\beta},\qquad 
\left\{\chi^\star(\varphi^\star)\right\}_{\alpha\beta} = -\,\left\{\chi(\varphi)
\right\}_{\alpha\beta}.\label{cc12}
\end{eqnarray}

\end{document}